# Ordering process and ferroelectricity in a spinel derived from $FeV_2O_4$

Q. Zhang, K. Singh, F. Guillou, C. Simon, Y. Breard, V. Caignaert and V. Hardy

Laboratoire CRISMAT, UMR 6508, CNRS ENSICAEN, 6 Boulevard du Maréchal Juin, F-14052 Caen Cedex 4, France

## Abstract

The spinel $FeV_2O_4$ is known to exhibit peculiar physical properties, which is generally ascribed to the unusual presence of two cations showing a pronounced interplay between spin, orbital and lattice degrees of freedom ($Fe^{2+}$ and $V^{3+}$ on the tetrahedral and octahedral sites, respectively). The present work reports on an experimental re-investigation of this material based on a broad combination of techniques, including x-ray diffraction, energy dispersive and Mössbauer spectroscopies, as well as magnetization, heat capacity, dielectric and polarization measurements. Special attention was firstly paid to establish the exact cationic composition of the investigated samples, which was found to be $Fe_{1.18}V_{1.82}O_4$. All the physical properties were found to point out a complex ordering process with a structural transition at $T_S = 138$ K, followed by two successive magnetostructural transitions at $T_{N1} = 111$ K and $T_{N2} = 56$ K. This latter transition marking the appearance of electric polarization, magnetization data were analysed in details to discuss the nature of the magnetic state at $T < T_{N2}$. An overall interpretation of the sequence of transitions was proposed, taking into account two spin couplings, as well as the Jahn-Teller effects and the mechanism of spin-orbit stabilization. Finally, the origin of ferroelectricity in $Fe_{1.18}V_{1.82}O_4$ is discussed on the basis of recent models.

# I. INTRODUCTION

Multiferroic materials have recently attracted increasing attention, both aimed at challenging fundamental properties and at exploiting them in multifunctional device applications.[1] There is a particular interest in the search for novel multiferroic materials with spontaneous magnetization and large polarization that persist up to high temperature, as well as a strong magnetoelectric coupling between magnetism and electric polarization.[2-3] The spinel oxide family[4] has various fascinating physical properties, but distinct from perovskite oxides[5], only few ferroelectrics were reported so far. The reasons why the spinel oxides do not favor either conventional ferroelectricity or magnetically driven ferroelectricity are still not clear. As for the latter class of ferroelectrics, one of the archetypical examples in spinels is $CoCr_2O_4$. This compound was reported to exhibit magnetically driven ferroelectricity associated to the onset of conical spin modulation (below 25 K), i.e., a behavior in accord with the spin-current model which can explain ferroelectricity in spiral magnet.[7] More recently, Giovannetti[8] investigated three vanadium spinels $AV_2O_4$ (A = Cd, Zn, and Mg), and found that the magnetically driven ferroelectricity exists only in $CdV_2O_4$. Experimental studies and *ab initio* calculations of the electric polarization suggest that the ferroelectricity in $CdV_2O_4$ –which exhibits a collinear magnetic structure– is driven by local exchange striction rather than the more common scenario of spiral magnetism.

Among spinel oxides, $FeV_2O_4$ is an interesting compound owing to a complex sequence of structural and magnetic phase transitions[9-11]: structural transitions at ~140 K (from cubic to tetragonal), at ~110 K (from tetragonal to orthorhombic) accompanied with the onset of collinear ferrimagnetism, and finally at ~ 70 K (from orthorhombic to tetragonal again), leading to a ground state whose magnetic nature is still obscure. The dielectric constant of $FeV_2O_4$ has been reported[12] to show a memory effect in small magnetic fields. The "double" distortions of $FeO_4$ tetrahedra and $VO_6$ octahedra make the physical properties of $FeV_2O_4$ very complex. In this context, it was of interest to investigate whether ferroelectricity could emerge in this peculiar spinel compound. These issues motivated us to re-investigate $FeV_2O_4$ by a combined use of various techniques, including Mössbauer spectroscopy, magnetization, ac susceptibility, heat capacity, dielectricity and ferroelectricity, and to address some features often neglected so far, such as the role of the exact cation distribution and the nature of the magnetic state below $T_{N2}$.

## II. EXPERIMENTAL DETAILS

The synthesis of $FeV_2O_4$ is made complex by the easy formation of secondary phases. In the present work, we designed a multiple-step solid state reaction and employed a different atmosphere from that used in previous reports[10, 13-14] to prepare polycrystalline samples. Only two starting materials of high-purity, $Fe_2O_3$ and $V_2O_3$, were mixed in the desired proportions and pressed into bars. The bars were heated firstly at 500 °C for 12 h, then at 800 °C for 24 h under a mixture of Ar and $H_2$ (1%) before being sintered at 1100 °C for 72 h under the same atmosphere. Finally, these ceramics were pulverized, mixed, pressed again into bars and heated at 1100 °C for 30 h again to homogenize the samples.

X-ray powder diffraction (XRPD) pattern was collected at room temperature over an angular range $5° < 2\theta < 120°$ with an X'pert Pro diffractometer working with the Cu K$\alpha$ radiations. Structural Rietveld refinement was performed using the FULLPROF program incorporated in the WinPLOTR Package.[15] Cationic homogeneity was checked by means of energy dispersive spectroscopy (EDS) realised on a TEM Jeol 2010.

The Mössbauer spectrum was registered at room temperature in transmission geometry using a constant acceleration spectrometer and a $^{57}$Co source diffused into a rhodium matrix. The spectrum was fitted with Lorentzian lines using the MOSFIT program[16] and the isomer shift values are given with respect to metallic iron at 293 K.

The temperature/field dependence of *dc* magnetization and the frequency dependence of the *ac* susceptibility in $h_{ac}$=10 Oe were recorded by means of a Physical Properties Measurement System (PPMS, Quantum Design). Heat capacity measurements were carried out in the same PPMS using a semiadiabatic relaxation method. Firstly, the whole temperature range was analysed with the built-in 2τ relaxation model. Secondly, we applied around each transition the Single Pulse Method (SPM), where C(T) is derived from a *point-by-point* analysis of the time response along *each branch* of a large heating pulse crossing the whole transition.[17] This second method yields better resolution in temperature and allows possible hysteresis to be detected.[18]

Complex dielectric permittivity was measured on parallel plate capacitor geometry with dimensions of 1.29 mm × 1.12 mm × 0.32 mm. Silver paste was used to make electrodes. Dielectric measurements were performed using Agilent 4284A LCR meter in different magnetic fields from zero to 140 kOe. The temperature and the magnetic field were controlled by the PPMS. Dielectric permittivity ε' and the losses tanδ were measured during cooling and warming (2 K/min) with a 1 V *ac* bias field at different frequencies (1 kHz to 100 kHz). Polarization was measured with a Keithley 6517 A electrometer in a coulomb mode with automatic current integration facility. Isothermal magnetodielectric effect was measured at different temperatures, and in fields between ±140 kOe using a sweep rate of 100 Oe/sec. In these measurements, the magnetic and electric fields were perpendicular to each other.

### III. RESULTS

#### III.1 Chemical and structural characterizations

The EDS analyses were performed on fifty crystallites. As two of them contained only iron, a second phase was introduced in the Rietveld refinement process; the corresponding set of very small peaks was totally indexed as metallic iron (SG Im-3m, $a$=2.8707Å; mass percentage content < 2 %). The other 48 crystallites revealed a good cationic homogeneity but their Fe/V ratio differs slightly from the expected one, leading to the approximate cationic formula $Fe_{1.15}V_{1.85}$. Taking into account this formula, the main phase was very well refined with the cubic F$d$-3$m$ space group ($a$= 8.4609 Å) and the usual atomic positions of the spinel structure[19] i.e., tetrahedral site: Fe (8 a: 1/8,1/8,1/8); octahedral site : V/Fe (16 d:1/2,1/2,1/2) and O (32 e: 0.26015, 0.26015, 0.26015). The difference pattern plot in Fig. 1 attests of the goodness of the fit ($\chi^2$=2.33 $R_{Bragg}$ = 4.23); note that an attempt to refine the main phase with the composition $FeV_2O_4$ was realized and leads to slightly worse results (($\chi^2$=2.33 $R_{Bragg}$ = 4.34).

Figure 2 shows the Mössbauer spectrum of our compound at room temperature. The Mössbauer spectrum is rather similar to the one obtained for $Fe_{1.2}V_{1.8}O_4$ by Abe et al.[20] and consists of two sharp Lorentzian lines with isomer shifts of 0.454 mm/s and 0.946 mm/s. These isomers shifts are characteristic of trivalent iron in octahedral coordination and divalent iron in tetrahedral coordination.[21] The two lines are in the ratio $Fe^{3+}/Fe^{2+}$=15/85. Assuming in a first approximation that vanadium is only trivalent and located on the octahedral sites, the cationic formula can thus be written as $(Fe^{2+})[Fe^{3+}_x V^{3+}_{2-x}]$, where parentheses and square brackets denote cation sites of tetrahedral (A sublattice) and octahedral (B sublattice) coordination, respectively. Accordingly, one can estimate x = 15/85 ≈ 0.18, leading to the formula $Fe_{1.18}V_{1.82}O_4$. Note that a Rietveld refinement of XRPD pattern was also performed with the composition $Fe_{1.18}V_{1.82}O_4$, which leads to the same reliability factors to the formula $Fe_{1.15}V_{1.85}O_4$.

### III.2 Magnetization and Heat capacity

The temperature dependence of the zero-field-cooled (ZFC) and field-cooled (FC) magnetization of $Fe_{1.18}V_{1.82}O_4$ in a moderate field of 5 kOe is shown in Fig. 3. With decreasing the temperature, a sharp increase in magnetization occurs below ~ 120K due to the paramagnetic (PM)-to-ferrimagnetic (FI) transition[8-10], while another anomaly is observed at a lower temperature on both the ZFC and FC curves. The $T_{N1}$ and $T_{N2}$, corresponding to the two peaks in the first derivative of the FC magnetization, are found to be 110 K and 56 K respectively, which accords basically with the previous reports in nominal $FeV_2O_4$.[9-11] There is negligible thermal hysteresis around $T_{N1}$, whereas a divergence between ZFC and FC curves emerges below around 90 K and becomes prominent at low temperature due to the anomalous decrease of ZFC magnetization. On the basis of *ac* susceptibility measurements, Nishihara[11] claimed the existence of a spin-glass-like state having a freezing temperature of 85.5 K, which could thus be responsible for that divergence between the ZFC and FC curves. To re-investigate this issue, the temperature dependences of both in-phase $\chi'$ and out-of-phase $\chi''$ signals of *ac* susceptibility were measured at frequencies of 100, 1000 and 10000 Hz. As shown in Fig. 4 (a), there are two anomalies at 110 K and 56 K in the $\chi'(T)$ curve, in accord with the two magnetic transitions observed in both ZFC and FC magnetization curves. Moreover, in contrast to Ref. 11, there is no peak between 80 and 100 K on the $\chi''(T)$ curve, precluding the existence of a spin-glass-like behavior in our compound.

Figure 5 shows the temperature dependence of FC magnetization curves in different magnetic fields. Both $T_{N1}$ and $T_{N2}$ are shifted to higher temperatures with the increase of the magnetic field, the effect being less pronounced on $T_{N2}$. In other respects, one observes on the FC curves a gradual increase in the magnetization below $T_{N2}$. In the inset of Fig. 5, the magnetic hysteresis loop at 5 K shows a hard magnetic behavior: The magnetization is not saturated up to 50 kOe and the coercive field is quite large, being ~15 kOe.

On the zero-field heat capacity curve, C(T), shown in Fig. 6(a), three peaks clearly emerge. For each of them, the temperature of the maximum, as well as that corresponding to the maximum negative slope, can be determined with an experimental uncertainty lower than 1 K. Referring to the maximum in C, the transitions are at ≈ 135 K, 109 K and 55 K. If one rather considers the maximum in $|dC/dT|$ –which is more appropriate in case of a second-order transition– these transitions are found to take place at ≈ 138 K, 111 K and 56 K. Comparing this calorimetric data with the magnetic results of the previous section and the structural characterizations reported by Katsufuji et al.,[9] these three temperatures can be ascribed to $T_S$, $T_{N1}$ and $T_{N2}$, respectively. It must be emphasized that no other anomalies were observed on the C(T) curve. In particular, there is neither indication of a transition around 4 K, as previously suggested on the basis of M(T) curves,[11] nor around 35 K, as observed in the structural characterization of single crystals[9].

In Fig. 6(a), the peak at the structural transition ($T_S$) is found to be quite broad, which is in agreement with the continuous evolution from cubic to tetragonal symmetries previously pointed out by Tanaka *et al.*.[22] The two lowest transitions at $T_{N1}$ and $T_{N2}$ are magnetostructural transitions, which could have been expected to present a first-order character. Nevertheless, no hysteresis could be detected when using our scanning method[17], and one observes that the peaks exhibit a lambda shape which is rather typical of second-order transitions. Actually, such a feature is consistent with the absence of discontinuity in the temperature evolution of the cell parameters around these transitions [9].

About the influence of the magnetic field, Fig 6 (b) shows a C/T vs. T plot focused on the temperature range of the transitions, for H=0 and H=90 kOe. As expected for a purely structural transition, one observes that the location of $T_S$ is almost unaffected by the field. Meanwhile, both magnetic transitions at $T_{N1}$ and $T_{N2}$ are broadened and shifted towards higher temperatures in 90 kOe. As expected, these effects are more pronounced at $T_{N1}$ which marks the onset of a ferrimagnetic order, while $T_{N2}$ only corresponds to an increase in the net magnetization as temperature is decreased.

### III. 3. Dielectric, polarization and magnetoelectric properties

The resistivity of $Fe_{1.18}V_{1.82}O_4$ is found to increase rapidly with the decrease of the temperature. At 150 K, the resistivity is as high as 2.6 $10^6$ $\Omega$cm suggesting that $Fe_{1.18}V_{1.82}O_4$ is a good insulator below this temperature. Fig. 7 (a) presents the temperature dependence of dielectric permittivity ε' at 100 kHz during warming process (2 K/min) without magnetic field, and Fig. 7(b) corresponds to the first derivative of $\varepsilon'$ (d$\varepsilon'$/dT) in different magnetic fields. At around 140 K, there is a kink in d$\varepsilon'$/dT (Fig. 7b) curve corresponding to $T_S$. Such a small anomaly reflecting the cubic-to-tetragonal transition is also observed in the $\varepsilon'$(T) curve of FeCr$_2$O$_4$ at a similar temperature, ≈ 140 K.[23] With further decrease of the temperature, two other anomalies are observed around the magnetic transition temperatures $T_{N1}$ (110 K) and $T_{N2}$ (56 K). These types of anomalies (a change in the slope) at magnetic/structural ordering are observed for Cr-based spinels ACr$_2$O$_4$ (A = Mn, Co, Ni and Fe)[23-24]. The locations of the two anomalies at lower temperatures are frequency independent (not shown here). It should be pointed out that the three anomalies at 140 K, 110 K and 56 K in dielectric measurements are well consistent with our magnetic and heat capacity results, implying strong correlation between charge, spin and lattice degrees of freedom.

To study the effect of magnetic field on the magnetic and structural transitions, dielectric measurements were measured in different constant magnetic fields (0 to 140 kOe). Fig. 7(b) demonstrates the effect of magnetic field on d$\varepsilon'$/dT plots at selected fields (0, 9 and 140 kOe). As expected, the kink in d$\varepsilon'$/dT at the structural transition temperature $T_S$ remains almost field independent, while the anomaly in d$\varepsilon'$/dT at the PM-FI transition temperature $T_{N1}$ become broader in high magnetic fields. As for the anomaly at the lowest temperature, the inset of Fig. 7(b) exhibits that the peak in d$\varepsilon'$/dT at $T_{N2}$ shifts towards higher temperatures with increasing the magnetic field, which supports a clear magnetodielectric coupling. The magnetodielectric couplings can also be observed in the variation of dielectric constant with respect to the magnetic fields [$\varepsilon'$(H) curve] at different temperatures. The representative curve at 5 K is shown in Fig. 7 (c). A clear magnetic-field hysteresis of ε' is observed, which is to some extent similar to the previous report about the ε'(H) curve of the nominal $FeV_2O_4$ performed at 20 K.[12] It is of interest to point that the coercive field $H_C$ in the ε'(H) curve is the same as observed in the M(H) curve at 5 K, comparing the same field excursion from -50 kOe to 50 kOe (see the inset of Fig. 5).

To check the presence of ferroelectricity in the magnetically ordered state, the temperature dependence of polarization was measured with an electrometer. A static electric poling field of ±60 kV/m was applied during cooling down to 8 K to align ferroelectric domains. Then, this poling field was removed and the stabilisation of polarization (P) vs time was recorded for 5000 sec at 8 K. The temperature dependence of polarization (P) was subsequently recorded during warming (5 K/min) in zero electric field and in different magnetic fields. Fig. 8 (a) evidences the presence of polarization only in the magnetic order state below $T_{N2}$. Moreover, reversing the poling electric field leads to a symmetric negative polarization. Note that the remnant polarization value (≈ 63 µC/m$^2$) is approximately 15 and 30 times higher than $CdV_2O_4$[6] and $CoCr_2O_4$[8], respectively. In addition, it is worthwhile noting that we have applied a relatively small electric poling field as compared with Ref. 6 and 8, and saturated P in $Fe_{1.18}V_{1.82}O_4$ should be higher at higher electric poling field in single domain state.

In the next step, we will demonstrate the clear evidence of magnetoelectric coupling. The sample was cooled down to 8 K with the same electric poling field (60 kV/m), before different magnetic fields (0-140 kOe) were applied to measure the temperature evolution of polarization. To avoid any effect of magnetic field history, the sample was systematically heated up to 300 K before each polarization vs. temperature curve. Fig. 8 (b) shows that, below $T_{N2}$, P decreases with increasing the magnetic field. These measurements are highly reproducible irrespective to the order of measurements in applied magnetic fields.

The inset presents the magnetic field variation of P at 20 K. The change in polarization ΔP ($P_{H=0}$ − $P_{H=140\ kOe}$) is 22 µC/m$^2$, which is the direct proof of strong magnetoelectric coupling existing in $Fe_{1.18}V_{1.82}O_4$. Therefore, we evidenced that one can effectively tune the polarization by magnetic field in $Fe_{1.18}V_{1.82}O_4$.

## IV. DISCUSSION

### IV.1/ The cation distribution

We found that the chemical and structural characterizations led to a composition $Fe_{1.18}V_{1.82}O_4$, which can even be written as $(Fe^{2+})[Fe^{3+}_{0.18}V^{3+}_{1.82}]$ if one assumes a perfect direct spinel. However, some authors suggested the possibility of a partial inversion, i.e., the presence of a small amount of $V^{3+}$ on the tetrahedral sites and $Fe^{2+}$ on the octahedral sites, leading for a stoichiometric $FeV_2O_4$ to a formulation $(Fe^{2+}_{1-\alpha}V^{3+}_{\alpha})[Fe^{2+}_{\alpha}V^{3+}_{2-\alpha}]$. For instance, Gupta et al. found α ≈ 0.1 from quantitative analysis of Mössbauer and x-ray diffraction spectra.[14] Combining the presence of an excess in iron with such an inversion, the general formulation that one has to consider is thus $(Fe^{2+}_{1-\alpha}\ V^{3+}_{\alpha})[Fe^{2+}_{\alpha}\ Fe^{3+}_{x}V^{3+}_{2-x-\alpha}]$, with x ≈ 0.18. The degree of inversion quantified by α is known to depend on the synthesis procedure. In our case, the synthesis involves a long-time plateau at 1100 °C, followed by cooling to room temperature at 1 °C/min. Calculations of the competition between $V^{3+}$ and $Fe^{2+}$ to be on the octahedral sites at 1100°C lead to α ≈ 0.2.[25] Owing to the slow cooling procedure that was adopted, one thus expects that the inversion parameter in our samples should lie within the interval $0 \leq \alpha \leq 0.2$.

It turns out that information about the actual α value can be derived from the magnetic properties. Let us consider the ferrimagnetic state taking place in the range $T_{N2} < T < T_{N1}$, which is widely admitted to be collinear.[10,12] In this regime, the magnetization recorded versus temperature or versus field are well consistent with each other, and perfectly reversible as soon as the field exceeds ~ 30 kOe. This indicates moderate anisotropy, which allows one to consider that the saturation magnetization derived from data recorded in a polycrystalline sample is representative of that of the collinear ferrimagnetic arrangement. In Fig. 5, one observes that, as the temperature is decreased, the M(T) in 140 kOe shows a first trend to saturation approaching $T_{N2}$ before undergoing an additional increase in M. The saturation magnetization of the collinear regime, referred to as $M_{sat}$(col.), can be approximated by a smooth extrapolation of the M($T_{N2} < T < T_{N1}$) curve down to lowest temperatures, leading to ≈ 2.15 $\mu_B$/fu. This experimental value must be compared to the theoretical one, i.e. $|M_A - M_B|$ where $M_A$ and $M_B$ are the saturation magnetizations of the A and B sublattices, respectively. For this purpose, one has to estimate the magnetic moment of each of the present cations. The simplest case is for tetrahedral $V^{3+}$ ($3d^2$) and octahedral $Fe^{3+}$ ($3d^5$), which are both expected to have a zero angular momentum, leading to 2 $\mu_B$ and 5 $\mu_B$, respectively. As for $Fe^{2+}$, small departures from the spin only value have been reported on both the tetrahedral and octahedral sites, corresponding to Landé factors (referred to as $g_A$ and $g_B$, respectively) that are comprised between 2 and 2.2.[26-28] The most complex issue deals with $V^{3+}$ in octahedral environment, a situation combining partial quenching of the orbital moment and a strong spin-orbit coupling. Inspection of the literature shows that the magnetic moment of $V^{3+}$ in octahedral sites ranges from 0.6 to 1.3 $\mu_B$. For instance, values between 0.61 and 0.65 $\mu_B$ have been found in $ZnV_2O_4$,[29,30] whereas $MnV_2O_4$ leads to values between 1.3 and 1.34 $\mu_B$.[31-33] Intermediate values close to 1.2 $\mu_B$ were observed in $CdV_2O_4$[34], as well as in other types of structures like $LaVO_3$ and $NaVO_2$.[35,36] As a starting point, we will thus consider that $0.6 \leq \mu(V^{3+}) \leq 1.3$. With the formula $(Fe_{1-\alpha}^{2+} V_\alpha^{3+})[Fe_\alpha^{2+} Fe_x^{3+} V_{2-x-\alpha}^{3+}]$, the saturation magnetization on the A and B sites are $M_A=2g_A(1-\alpha)+2\alpha$ and $M_B=2g_B\alpha+5x+(2-x-\alpha)\mu(V^{3+})$. Using x=0.18, it turns out that, *whatever the possible values of* $g_A$, $g_B$ and $\mu(V^{3+})$, finding $|M_A-M_B|$ close to 2.15 $\mu_B$/fu requires that $\alpha \approx 0$. Note that such a result is consistent with two Mössbauer investigations which reported negligibly small inversion parameter as long as x remains lower than 0.35[13] or than 0.66.[37] In the rest of the paper, we will thus consider that the cation distribution of our compound is $(Fe^{2+})[Fe_{0.18}^{3+}V_{1.82}^{3+}]$.

### IV.2/ The sequence of transitions

Let us now address the interpretation of the successive transitions found at $T_S$=138 K, $T_{N1}$=111 K and $T_{N2}$=56 K. When combining our data with the structural characterizations carried out by Katsufuji *et al.* on ceramic samples,[9] our series of transitions can be described as follows: as temperature is decreased, there is first a purely structural transition from cubic to high-temperature (HT) tetragonal (c<a) at $T_S$=138 K; then, a magnetostructural transition towards *Néel* collinear ferrrimagnetism takes place at $T_{N1}$=110 K, accompanied by a symmetry change from HT tetragonal to orthorhombic; finally, $T_{N2}$=56 K corresponds to a transition towards reversed low-temperature (LT) tetragonality (c>a), which is accompanied by a modification in the ferrimagnetic state.

At this stage, one must address a possible role of the iron excess in the physics of $Fe_{1+x}V_{2-x}O_4$. It turns out that most of the previous studies on "$FeV_2O_4$" were actually carried out on samples showing a non-zero x value. For instance, Katsufuji et al.[9] reported that 7% of the octahedral sites are occupied by Fe in their single crystals. We note this is close to the stoichiometry of our own samples (0.18/2=0.09). As detailed by Jeannot et al.[38] there are basic thermodynamical reasons which make very delicate the stabilization of stoichiometric $FeV_2O_4$. It turns out, however, that Rogers et al.[39] claimed they were able to obtain a precise control of x in the whole range $0 \leq x \leq 1$. Strikingly, it was found that the high-temperature structural transition for x=0 is from cubic to tetragonal with c>a ($T_S \sim 127$ K), whereas the behaviour presently observed with a first transition towards a tetragonal phase having c<a ($\sim 135$ K), followed by a second transition towards orthorhombic symmetry ($\sim 110$ K), *only takes place in a restricted x range approximately between 0.05 and 0.15.*

In Ref 9, the peculiar series of transitions seen in "$FeV_2O_4$" is ascribed to the coexistence of tetrahedral ($Fe^{2+}$) and octahedral [$V^{3+}$], which are both *orbitally active* cations. Katsufuji et al.[9] proposed a microscopic scenario for the transition taking place at $T_N$ (i.e., our $T_{N1}$) between the paramagnetic tetragonal structure (c<a) and the ferrimagnetic orthorhombic structure. In their scenario, the main role of [$V^{3+}$] is to trigger a reorientation of the ($Fe^{2+}$) spins below the collinear ferrimagnetic transition. We suggest that the spin and orbital behaviours associated to [$V^{3+}$] could play in themselves a still more decisive role, leading us to address the following issues: (i) the nature of the magnetic ordering at $T < T_{N2}$; (ii) the dual role of [$V^{3+}$], resulting from the presence of empty $e_g$ orbitals and from its sensitivity to a mechanism of spin-orbit (SO) stabilization.

### IV.3/ Magnetic state at T < $T_{N2}$

The FC M(T) curves in large magnetic fields exhibit an increase in magnetization as temperature is decreased below $T_{N2}$, indicating the onset of a ferrimagnetic arrangement having a saturation magnetization larger than the one of the collinear phase. This magnetic order at T < $T_{N2}$ also exhibits a pronounced magnetic anisotropy, as attested to by the shape of the ZFC M(T) curves (see example on Fig. 3). As suggested since the 1960s[10,39,40], this transition at $T_{N2}$ is most likely related to a modification in the ferrimagnetic ordering induced by the magnetic coupling between the spins on the octahedral sites, $J_{BB}$. In the spinel vanadates $AV_2O_4$, there is indeed a significant antiferromagnetic interaction between the spins of [$V^{3+}$] that always come into play in the ground state of these materials. Starting from a collinear *Néel* ferrimagnetism induced by the larger coupling between the A and B sublattices ($J_{AB}$), this secondary antiferromagnetic interaction is expected to induce some canting within the $V^{3+}$ spins on the B sites. More specifically, it turns out that these spins often exhibit a triangular arrangement of Yafet-Kittel type, as found for instance in $MnV_2O_4$.[41] Strikingly, it can be noted that the size of the magnetization step below $T_{N2}$ (see Fig. 5) is similar to that taking place in $MnV_2O_4$ at nearly the same temperature. In the triangular ferrimagnetic state of $MnV_2O_4$, several studies showed that the [$V^{3+}$] spins are oriented along two symmetrical orientations, each tilted by an angle θ close to 65° from the direction of the ($Mn^{2+}$) spins, i.e., the *c* axis.[31,32,33] To estimate such an angle in the case of $Fe_{1.18}V_{1.82}O_4$, one can consider the height of the magnetization step at the canting transition. In the collinear ferrimagnetic state, we noted that $M_{sat}$(col.) = $2g_A$-0.9-1.82μ($V^{3+}$). Using the experimental $M_{sat}$(col.) ≈ 2.15 μ$_B$/fu and the mean value of the allowed range for $g_A$ (i.e., 2.1), one obtains $\mu(V^{3+}) \approx 0.63\ \mu_B$. In the triangular configuration, $M_{sat}$ is expected to be increased to $M_{sat}(tri.) = |M_A - M_B \cos(\theta)|$. The above values of $g_A$ and $\mu(V^{3+})$ lead to $\cos(\theta) = [4.2 - M_{sat}(tri.)]/2.05$. Our data indicates that M below $T_{N2}$ tends to saturate towards 2.34 $\mu_B$/fu. Nevertheless, such data recorded on polycrystalline samples can substantially underestimate $M_{sat}(tri.)$ owing to the large magnetic anisotropy present at T < $T_{N2}$.[12] Therefore, one must rather consider data recorded on single crystals with the field applied along the easy axis, i.e., B//c, which leads to $M_{sat}(tri.) \approx 3.2\ \mu_B$/f.u.[9,12], and thus $\theta \approx 60°$. We emphasize the value of this canting angle is just an estimate, since its derivation combines several experimental uncertainties.

### IV.4/ Interplay between the spin and orbital physics of $(Fe^{2+})$ and $[V^{3+}]$

Let us first present the four mechanisms that we assume to predominantly drive the structural and magnetic transitions in $Fe_{1.18}V_{1.82}O_4$. In terms of spin-spin interactions, the two main couplings are $J_{AB}$ and $J_{BB}$ which are both antiferromagnetic, the latter being smaller than the former. In terms of orbital-lattice couplings, the peculiarity is that both $(Fe^{2+})$ ($e_g^3 t_{2g}^3$) and $[V^{3+}]$ ($t_{2g}^2 e_g^0$) are orbitally active cations.[42]

For $(Fe^{2+})$, one expects a Jahn-Teller (JT) effect, splitting the $e_g$ doublet to host the sixth electron in the lowest singlet. Either elongation or compression of the FeO$_4$ tetrahedron can lift this degeneracy, leading to $d_{z2}$ and $d_{x2-y2}$ singlets. Goodenough showed that the type of distortion adopted by $(Fe^{2+})$ actually depends on the nature of the cations present at the B sites.[43] First, he emphasized that there is no cooperative JT effect in the absence free 3d orbitals directed toward the nearest-neighbor anions, as exemplified by FeAl$_2$O$_4$.[37] In contrast, the presence of B site cations having empty $e_g$ orbitals tends to stabilize a compression of FeO$_4$, leading to tetragonality with c<a, as illustrated by FeCr$_2$O$_4$.[44] In other respects, the presence of magnetostrictive cations on the B sites can trigger a competitive mechanism of spin-orbit (SO) stabilization that favors an elongation of FeO$_4$, as it is the case for instance with [Fe$^{2+}$] in Fe$_2$TiO$_4$.[27,45] In Fe$_{3-x}$Cr$_x$O$_4$, there is a competition between the two above mechanisms, leading to tetragonal phases with either c<a or c>a, as well as an orthorhombic intermediate regime.[43,46]

As for [V$^{3+}$], let us first mention that their empty e$_g$ orbitals can favor a compression of the FeO$_4$, as it is the case with [Cr$^{3+}$]. Moreover, [V$^{3+}$] can undergo either JT or SO stabilizations of the t$_{2g}$ shell, leading to distortions of opposite signs.[42] It was noticed that the JT effect alone would favor a splitting of the t$_{2g}$ triplet lowering the (d$_{zx}$,d$_{yz}$) doublet to host the two 3d electrons, i.e., an elongation of the VO$_6$ octahedron.[43] In contrast, the SO stabilization rather requires to place one electron into a doublet[42] thus favoring the compression of VO$_6$, which is actually the distortion most commonly observed in vanadates.[47,48] In turn, we previously noted that the onset of a cooperative orbital distortion associated to such a SO stabilization at the octahedral sites favors an elongation of the tetrahedra.[43] Accordingly, this mechanism can be claimed to be at the origin of the cubic-to-tetragonal transition with c>a, that was reported in stoichiometric FeV$_2$O$_4$.[39] In their study of Fe$_{1+x}$V$_{2-x}$O$_4$, Rogers *et al.* also showed that x as small as 0.05 can strongly modify the structural transition.[39] It must be emphasized that orbital ordering on the B sites of spinels can be easily destabilized by substitutions with cations being orbitally "neutral", as it was shown for instance by Adachi *et al.* in $(Mn^{2+})[V^{3+}_{2-x}Al^{3+}_x]$.[49] In $(Fe^{2+})[Fe^{3+}_{0.18}V^{3+}_{1.82}]$, it turns out that the [Fe$^{3+}$] are also neutral from an orbital viewpoint since they have L = 0. Their presence on the B sublattice can thus act as a *random field*[49] impeding the onset of a collective orbital distortion related to SO stabilization.

Let us now propose a scenario for the series of transitions presently found in $(Fe^{2+})[Fe^{3+}_{0.18}V^{3+}_{1.82}]$. As temperature is decreased, the highest energy term that comes into play is the JT effect at the ($Fe^{2+}$) sites. The sign of the FeO$_4$ distortion is determined by the competition between the two mechanisms described above: (1) a collective SO stabilization of the [$V^{3+}$], which favors elongation of the tetrahedra; (2) the presence of free e$_g$ orbitals in [$V^{3+}$], which favors compression of the tetrahedra. Since Katsufuji *et al.* experimentally observed a compression of FeO$_4$ in a composition close to ours,[9] we suggest that the role of "orbital dilution" played by the Fe$^{3+}$ in $(Fe^{2+})[Fe^{3+}_{0.18}V^{3+}_{1.82}]$, can substantially weaken mechanism (1) in such a way that mechanism (2) becomes predominant at the cubic-to-tetragonal transition. Therefore, this leads to a tetragonal structure with c < a below $T_S$ = 138 K, in a way very similar to what is found in FeCr$_2$O$_4$ at 135 K. We suggest that the second energy term entering into play is the J$_{AB}$ interaction, which yields antiparallel alignment of the (Fe$^{2+}$) and [V$^{3+}$] spins when decreasing the temperature below $T_{N1}$. It turns out that the onset of this magnetic transition has a considerable impact on the lattice distortions. Indeed, the strength of the SO stabilization is expected to be enhanced in a magnetically ordered regime[47,48], a feature which in our case can assist the onset of a collective SO stabilization of the [V$^{3+}$]. Note that a strengthening of the SO mechanism is experimentally supported by an enhancement of the VO$_6$ compression below T$_{N1}$.[9] Owing to the elastic energy associated with a cooperative orbital ordering at the B sites, mechanism (1) can override mechanism (2), inducing a reversal of the FeO$_4$ distortion, as it is observed in the synchrotron X-ray powder diffraction data of Ref. 9. Then, as the temperature is further decreased, the $J_{BB}$ coupling is proposed to induce a canting within the [$V^{3+}$] spins at $T_{N2}$, as previously discussed. We note that such a "non-collinear" spin arrangement was claimed to reinforce the SO mechanism, [39] which would be qualitatively consistent with the stabilization of the distortion parameters that is observed below $T_{N2}$[9].

In terms of global symmetry of the structure, the regime $T_{N1} < T < T_S$ is mainly influenced by the compression of FeO$_4$, which thus results in tetragonal symmetry with c < a. For T < $T_{N2}$, one gets elongated FeO$_4$ coexisting with compressed VO$_6$, which leads to a tetragonal phase with c>a, as in the stoichiometric FeV$_2$O$_4$.[39] As for the orthorhombic symmetry found in the range $T_{N2}$ < T < $T_{N1}$, it can be regarded as a necessary intermediate regime between the two types of tetragonality.[9] Our suggestion about the sequence of structural and magnetic transitions is summarized in Fig. 9, together with the types of FeO$_4$ and VO$_6$ distortions that were observed in Ref. 9.

### IV.5/ The origin of the polarization

Polarization in Fe$_{1.18}$V$_{1.82}$O$_4$ emerges below $T_{N2}$ at the orthorhombic to tetragonal (c>a) transition. The low temperature crystal structure (*I4$_1$/amd*) is centrosymmetric and thus should not allow standard polarization. We also note that the remnant polarization (63 µC/m$^2$) is much higher than in CoCr$_2$O$_4$[6] or CdV$_2$O$_4$[8] spinels, and is comparable to that of the so-called improper ferroelectrics[1, 50].

Two possible models[51] are usually proposed to explain the origin of such magnetically induced polarization. The first one is based on exchange striction (the magnetic exchange coupling J depends on the orbital occupancy and thereby on the shape and orientation of these orbitals) and may correspond to collinear magnetic structures.

This model is used to explain the origin of polarization in collinear magnetic states e.g. $YMn_2O_5$[52] and $Ca_3(Co,Mn)_2O_6$[53]. Recently, Giovannetti *et al.* observed polarization in $CdV_2O_4$ in a collinear magnetic state and explained it on the basis of exchange striction[8]. The presence of polarization in collinear magnetic state is also observed in spinel chromites (polycrystalline) $FeCr_2O_4$[23]. Since in our case, there is no polarization in the collinear phase, it suggests that this scenario may not be applicable.

The second model is based on reverse Dzyaloshinskii-Moriya interaction and is relevant to spiral magnetic structures (spin current model)[7]. The origin of polarization in the noncollinear magnetic structure of various oxides, such as $CoCr_2O_4$[6], $RMnO_3$ (R=Tb and Dy) [50, 54-55], $Ni_3V_2O_8$[56], $MnWO_4$[57] and $Mn_{0.9}Co_{0.1}WO_4$[58], and $LiCuVO_4$[59] can be explained by this spin-current model. This model is started from the spin-orbit interaction term of the electronic Hamiltonian, and the electric polarization $P_{ij}$ induced between the canted magnetic moments ($S_i$ and $S_j$) on the neighbouring sites $i$ and $j$ is given as:

$$P_{ij} = a\, e_{ij} \times (S_i \times S_j). \qquad (1)$$

Here, $e_{ij}$ denotes the unit vector along the direction from the spin site $i$ to $j$ and $a$ is the proportional constant depending on the spin exchange interaction and the spin-orbit interaction.

Recently, Kaplan[60] further developed the conventional spin-current model by adding previously omitted terms in the existing microscopic models as the additional contributions to the polarization, and the formula (1) in the conventional spin-current model[7] is regarded as only one form of polarization in this extended spin current model. It is therefore demonstrated that a pair of magnetic atoms with canted spins $S_a$ and $S_b$ can give rise to an electric dipole moment *P*, which is named "canted-spin-caused electric dipoles".

As it is explained in the magnetic section, $T_{N2}$ probably corresponds to a magnetic transformation from collinear to triangular magnetic states. Since polarization appears only below $T_{N2}$ and is tunable with magnetic field, this suggests that the spin canting is responsible for the observed polarization. Thus, it appears that the observed polarization in $Fe_{1.18}V_{1.82}O_4$ could be explained by the spin current model. Note that such a spin-canting-induced ferroelectricity (with a large $P \approx 100 \ \mu C/m^2$) was recently claimed to take place in the canted antiferromagnet $SmFeO_3$[61].

If one tries to calculate the corresponding atomic displacements, they are very small and probably very difficult to measure directly by diffraction. As a consequence, this magnetically induced "ferroelectricity" needs accurate knowledge of the magnetic symmetry, which should be in this compound strongly linked to the orbital occupancy ordering. Unfortunately, to the best of our knowledge, there is no information available regarding the magnetic group of $Fe_{1-x}V_{2-x}O_4$ spinels.

## V. CONCLUSION

In summary, it was found that the spinel $Fe_{1.18}V_{1.82}O_4$ is a new multiferroic material, which exhibits spontaneous magnetization and polarization. A PM-FI transition occurs at $T_{N1}$ (=110 K), while another complex magnetic transition, probably from collinear *Néel* configuration to a triangular ferrimagnetic structure, emerges at $T_{N2}$ (=56 K). This latter transition is marked by a step in magnetization, a peak in heat capacity, an anomaly in the dielectric constant, and the appearance of polarization. It was found that the application of a magnetic field shifts all these signatures associated to $T_{N2}$ to higher temperatures, while it also clearly affects the value of the polarization, revealing a significant magnetoelectric coupling. It is suggested that the presence of canted spins in the triangular structure below $T_{N2}$ could be responsible for the appearance of ferroelectricity. The results of the present study strongly reinforce the need of neutron diffraction investigations in spinels derived from $FeV_2O_4$, in order to directly determine their magnetic structure below $T_{N2}$.

**Acknowledgments**

This work has been supported by the European project "SOPRANO" under Marie Curie actions (Grant. No. PITN-GA- 2008-214040) and French project "PR Refrigeration Magnétique".

 

Figure captions:

Fig. 1. Experimental (crosses), calculated and difference (solid lines) X-ray powder diffraction pattern of nominal $Fe_{1.15}V_{1.85}O_4$ at the end of refinement. The vertical bars are the Bragg positions for the phases $Fe_{1.15}V_{1.85}O_4$ (up) and metallic Fe (down).

Fig. 2. Mössbauer spectrum at room temperature, with fitting discussed in the text

Fig. 3. The temperature dependences of ZFC (triangles) and FCC (diamonds) magnetization curves in a moderate field of 5 kOe for $Fe_{1.18}V_{1.82}O_4$.

Fig. 4 The temperature dependences of (a) in-phase $\chi'$ and (b) out-of-phase $\chi''$ signals in *ac* susceptibility measured under different frequencies for $Fe_{1.18}V_{1.82}O_4$. The inset of Fig. 4 (a) shows the enlarged view of in-phase $\chi'$ around $T_{N2}$.

Fig. 5 The temperature dependences of FC magnetization in various fields from 100 Oe up to 140 kOe in $Fe_{1.18}V_{1.82}O_4$. The inset shows the magnetic hysteresis loop at 5 K.

Fig. 6 (a): Temperature dependence of the heat capacity in zero field for $Fe_{1.18}V_{1.82}O_4$, with dashed lines marking the three transitions; (b): Zoomed view of the C/T vs T curves in 0 and 90 kOe

Fig. 7 (a) Temperature dependence of the dielectric permittivity of $Fe_{1.18}V_{1.82}O_4$ measured at 100 kHz in zero magnetic field. (b) $d\varepsilon'/dT$ vs T curves in different magnetic fields, at 100 kHz. The inset shows the field-induced shift of the dielectric anomaly associated to $T_{N2}$. (c) Isothermal magnetodielectric effect at 5 K during increasing and decreasing magnetic field (100 Oe/sec) in $Fe_{1.18}V_{1.82}O_4$. The arrows mark the direction (increasing and decreasing) of magnetic field.

Fig. 8 (a): Temperature dependences of the zero-electric-field polarization in $Fe_{1.18}V_{1.82}O_4$, recorded during warming (5 K/min) after using a static electric poling field of +60 or -60 kV/m (see text for more details). (b): Temperature dependence of the polarization in different magnetic fields (same poling field of +60 kV/m). The inset shows the influence of the magnetic field on the polarization at T=20 K.

Fig. 9 Schematic illustration of our proposal for the origin of the sequence of transitions observed in $Fe_{1.18}V_{1.82}O_4$. The abbreviations PM and FI are for paramagnetic and ferrimagnetic, respectively. Our comments about the nature of the distortions of the $FeO_4$ tetrahedra and $VO_6$ octahedra are derived from the experimental results of Ref. 9.

Fig. 1.

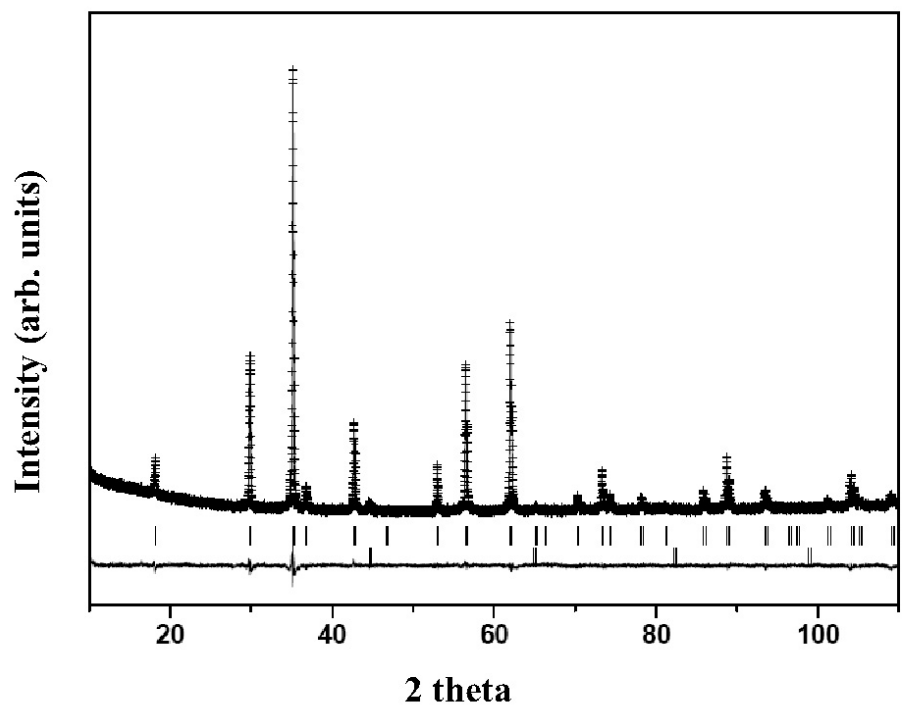

Fig. 2.

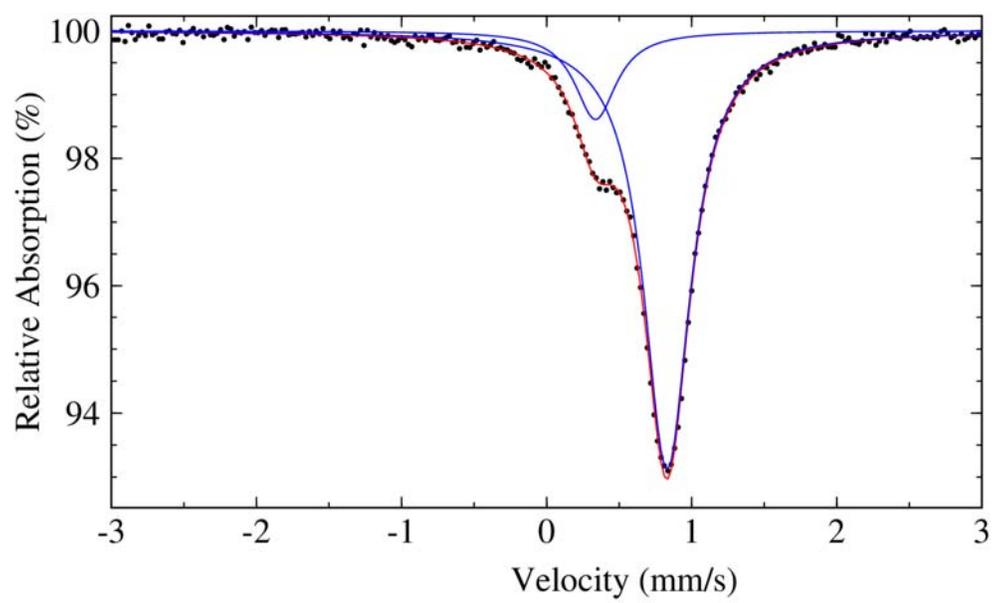

Fig. 3.

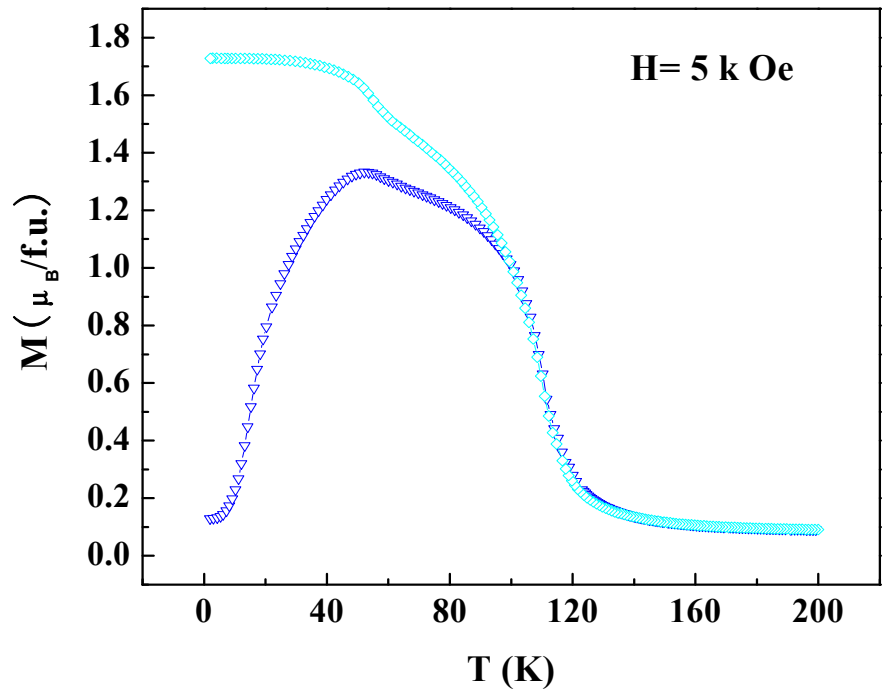

Fig. 4.

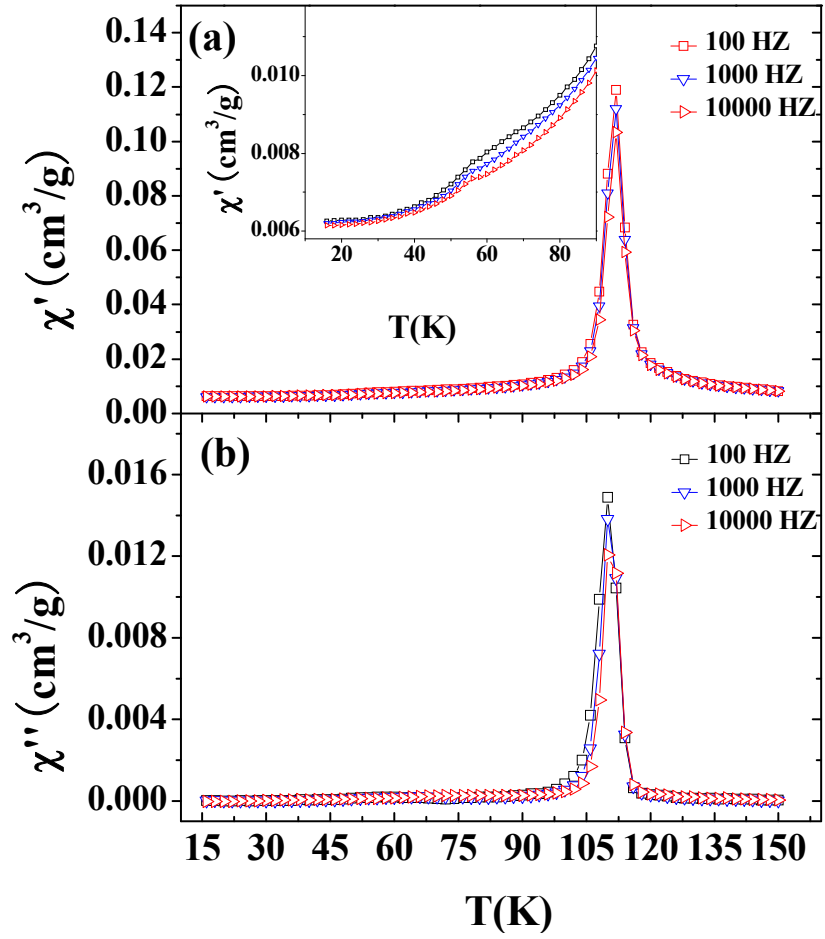

Fig. 5.

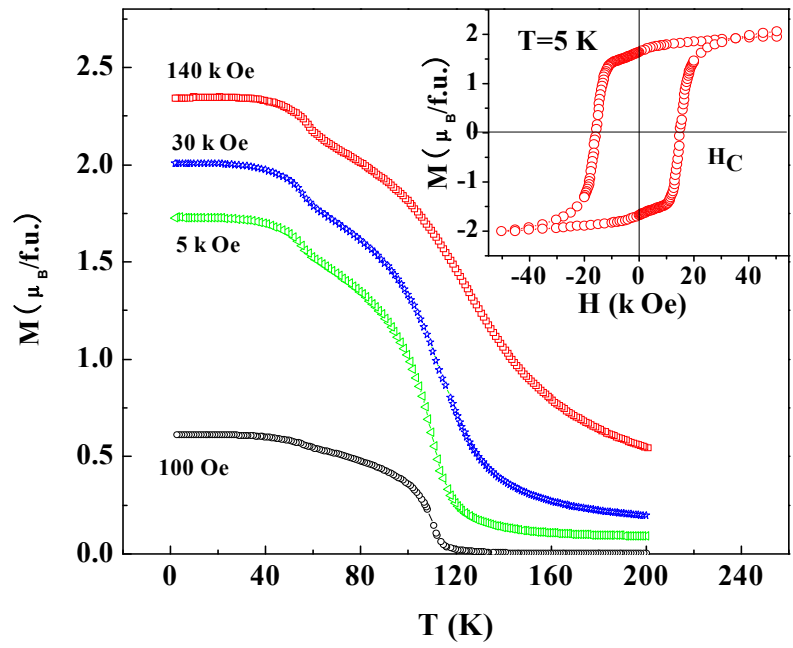

Fig. 6.

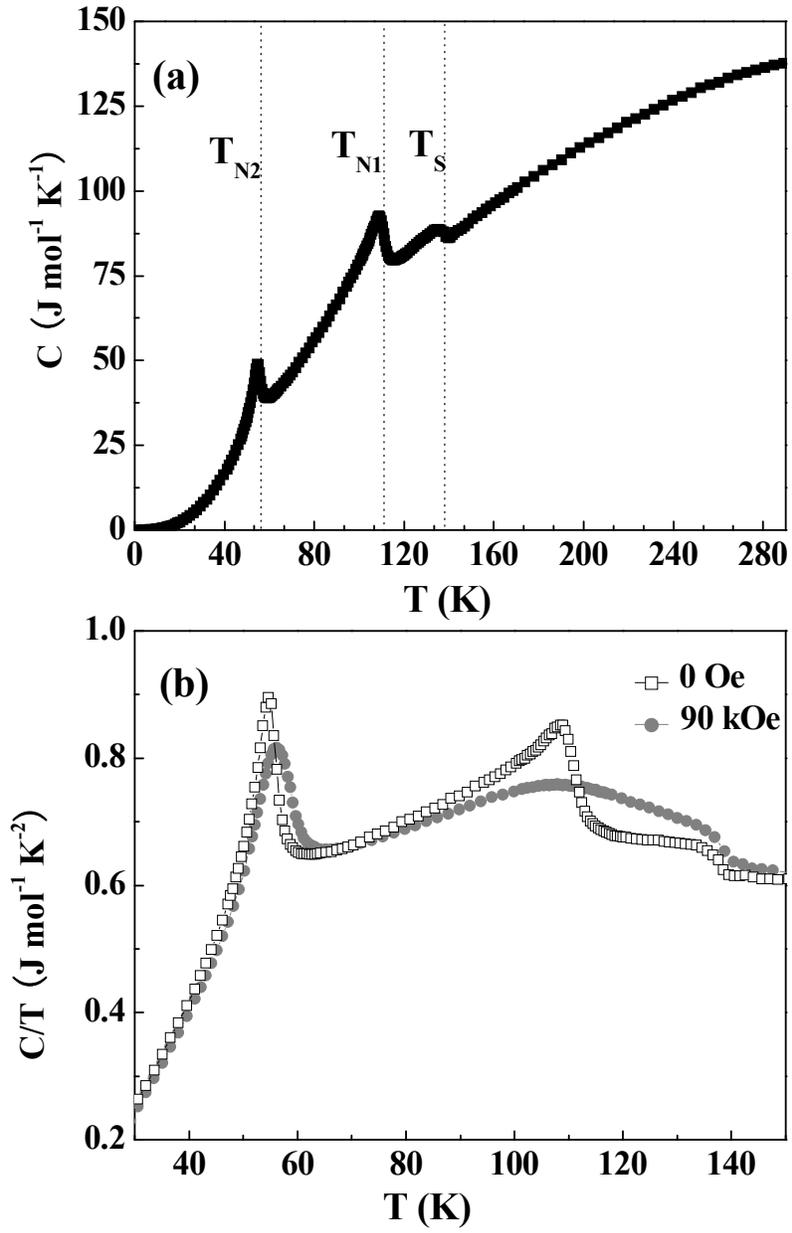

Fig. 7.

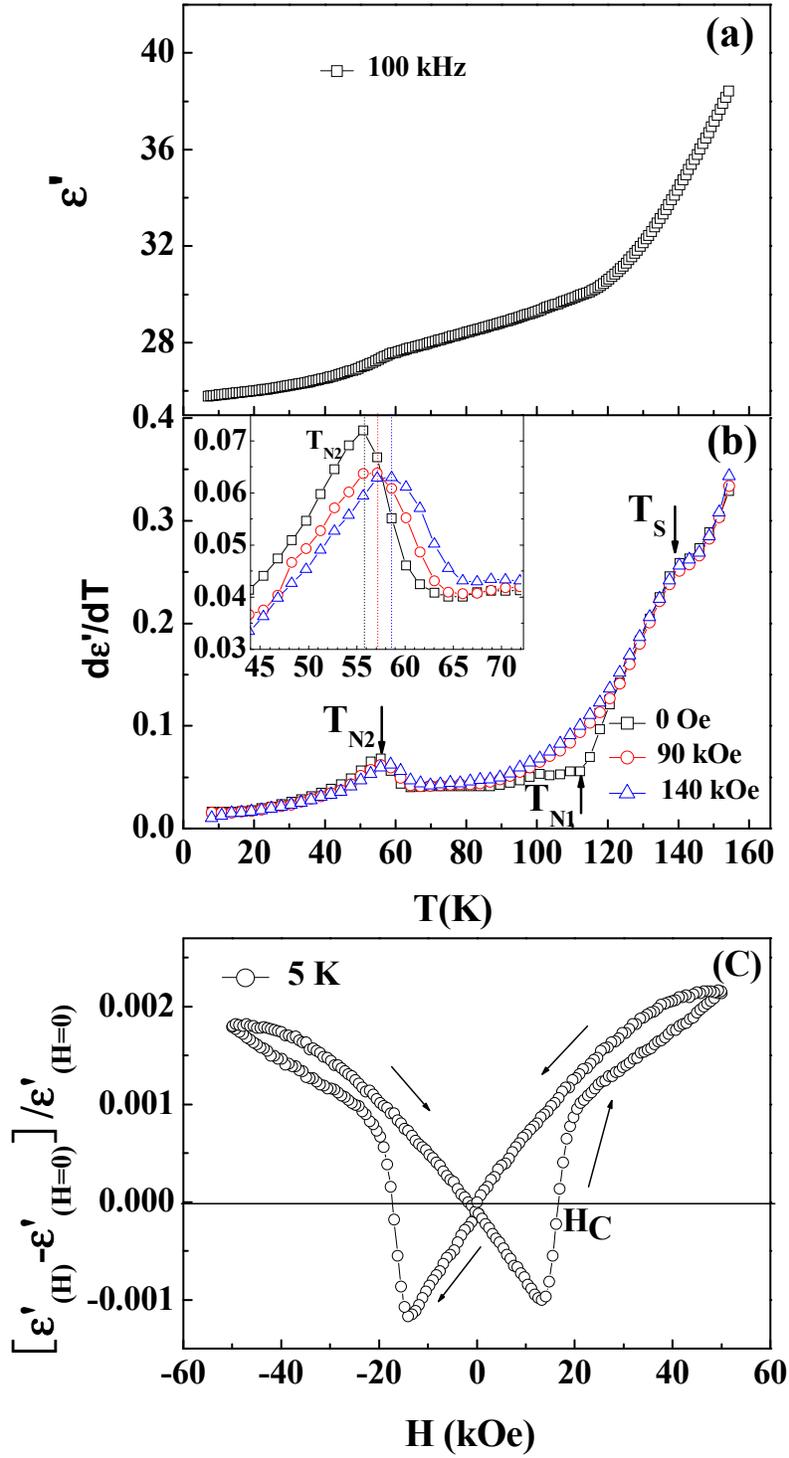

Fig. 8.

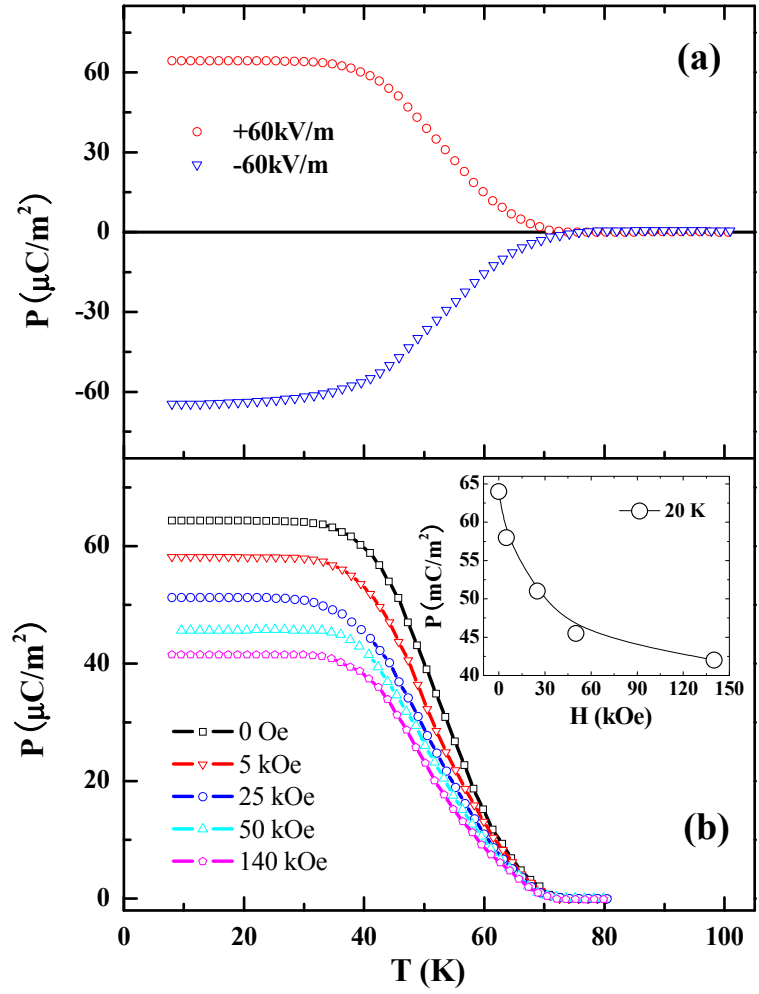

Fig. 9.

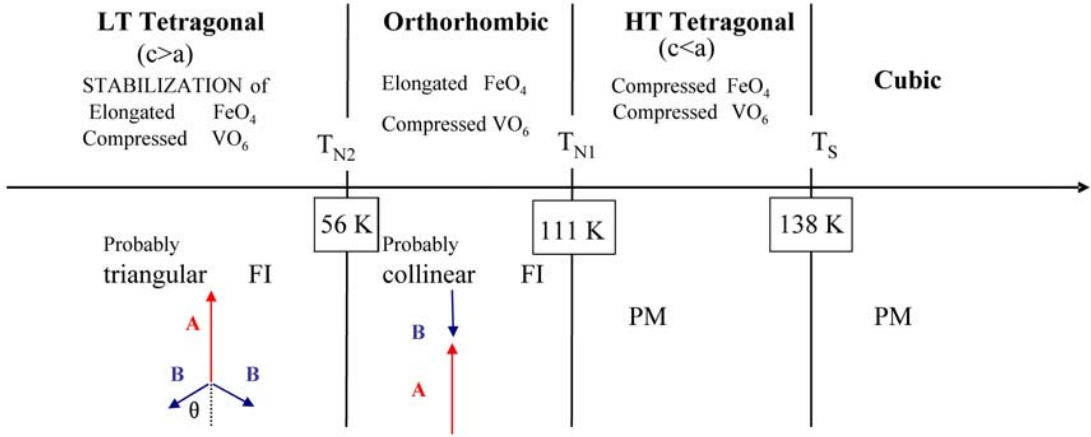